\documentclass[twocolumn,nofootinbib,preprintnumbers]{revtex4}

\usepackage{amsmath}
\usepackage{graphicx}

\newcommand{\nn}{\nonumber}
\newcommand{\ord}{\mathcal{O}}
\newcommand{\df}{\mathrm{d}}
\newcommand{\GeV}{\mathrm{GeV}}
\newcommand{\pythia}{\texttt{Pythia}}
\newcommand{\sherpa}{\texttt{Sherpa}}
\newcommand{\mo}{0}
\newcommand{\cut}{\mathrm{cut}}
\newcommand{\old}{\mathrm{old}}
\newcommand{\new}{\mathrm{new}}

\newcommand{\mx}{\mathrm{max}}
\newcommand{\ini}{\mathrm{ini}}
\newcommand{\PS}{\mathrm{PS}}
\newcommand{\ME}{\mathrm{ME}}

\newcommand{\cP}{\mathcal{P}}
\newcommand{\bP}{\bar{P}}

\begin{document}

\title{Gaining analytic control of parton showers}

\author{Christian W.~Bauer}
\author{Frank J.~Tackmann}
\affiliation{Ernest Orlando Lawrence Berkeley National Laboratory, University of California, Berkeley, CA 94720}

\begin{abstract}

Parton showers are widely used to generate fully exclusive final states needed to compare theoretical models to experimental observations. While, in general, parton showers give a good description of the experimental data, the precise functional form of the probability distribution underlying the event generation is generally not known. The reason is that realistic parton showers are required to conserve four-momentum at each vertex. In this paper we investigate in detail how four-momentum conservation is enforced in a standard parton shower and why this destroys the analytic control of the probability distribution. We show how to modify a parton shower algorithm such that it conserves four-momentum at each vertex, but for which the full analytic form of the probability distribution is known. We then comment how this analytic control can be used to match matrix element calculations with parton showers, and to estimate effects of power corrections and other uncertainties in parton showers.

\end{abstract}

\maketitle

%%%%%%%%%%%%%%%%%%%%%%%%%%%%%%%%%%%%%%%%%%%%%%%%%%%%%%%%%%%%%%%%%%%%%%%%%%%%%%%%
\section{Introduction}
%%%%%%%%%%%%%%%%%%%%%%%%%%%%%%%%%%%%%%%%%%%%%%%%%%%%%%%%%%%%%%%%%%%%%%%%%%%%%%%%

To analyze experimental data at high energy colliders, one needs precise theoretical predictions to compare measurements against. For such comparisons, Monte Carlo event generators that simulate fully exclusive events are indispensable tools~\cite{Skands:2005hd}. While it is possible to calculate simple observable distributions analytically, in most cases a direct comparison of such calculations with the data is very difficult, because the experimental analyses have to impose a variety of cuts, and detector efficiencies are, in general, not uniform over the measured distribution.

Event generators typically simulate events in three separate steps: First, a matrix element generator generates an event with few final-state partons, based on full matrix element calculations, which include all interference effects of the quantum field theory. Second, a parton shower adds additional partons using classical splitting probabilities. Finally, all partons are turned into hadrons according to some QCD model of hadronization.

In this paper we are concerned with the second step in an event generator, the parton shower. In practice it is impossible to generate all possible partonic final states using matrix element calculations, because the number of partons in the final state quickly becomes too large. In contrast, parton showers are based on splitting functions which describe the classical probability that a given mother parton splits into two daughter partons. Thus, with a certain probability described by the splitting functions, the parton shower turns a final state containing $n$ partons into one with $n+1$ partons. The advantage of using splitting functions over full matrix elements is that this procedure can be iterated to take a simple final state with a small number of partons and produce many additional partons using a Markov Chain.

The splitting functions describe the splitting in the collinear limit, where the mother and two daughters have large energy and small invariant mass. The parton shower also resums the leading Sudakov double logarithms. Recently it was shown~\cite{Bauer:2006qp} that the parton shower is formally reproduced as the leading order in an effective field theory expansion using soft-collinear effective theory~\cite{SCET}, which provides, in principle, a framework for systematic improvements. For example, one could attempt to sum large logarithms beyond the leading order or study power corrections. However, almost all theoretical improvements will necessarily go beyond the level of classical splitting probabilities. It is thus very unlikely that they can be incorporated into a Markov Chain algorithm. A typical example are matrix element calculations, for which it is hard to directly distribute events according to, and which must be explicitly matched with parton showers to avoid double counting~\cite{MLM,CKKW,Lonnblad:2001iq}.

Other important aspects are theoretical uncertainties in the generated distribution~\cite{Stephens:2007ah}. They arise from input parameters, like $\alpha_s$ and quark masses, as well as from higher-order power and perturbative corrections. Having a reliable estimate of these uncertainties becomes crucial at the LHC in searches for New Physics signals where the Standard Model background can only be estimated using Monte Carlo generators.

Related to these issues, important practical considerations must be taken into account as well. The three steps described above produce a theoretical distribution of events. However, the obtained events still have to be run through a detector simulation in order to compare them directly with experimental data. It is this additional step which consumes by far the most computing time in practice. Generating a typical event at the LHC before the detector simulation takes only a fraction of a second per CPU, whereas propagating an event through the detector simulation takes several minutes. In addition, it takes of the order of a megabyte to store a fully simulated event. The available amount of processing power and disk space thus leads to considerable practical limitations. For example, it is extremely impractical to resimulate the full event set each time the theory makes progress, or to simulate separate event sets using different parameter values in order to estimate theoretical uncertainties.

All of these problems can be solved if the exact distribution of the generated events is known. If this is the case, one can reweight the generated events according to a different theoretical distribution, even after the time-consuming detector simulation. This makes it possible to reuse existing simulated events, and thus allows one with small effort to study theoretical uncertainties and to include theoretical improvements whenever they become available.

For this reweighting approach to be possible, one has to have control of the precise functional form of the probability distribution underlying the event generator. For the matrix element generator, this is always the case by construction. However, it also requires one to have analytic control of the parton shower, which is generally not the case for the currently used parton showers. The reason is that a realistic parton shower needs to enforce four-momentum conservation at each vertex. While this is strictly speaking a subleading effect, it typically generates cross correlations between different splittings. Hence, although the basic probability distribution governing a single splitting can be obtained analytically, the analytic control over the full distribution is lost in the standard parton shower algorithms because of the way four-momentum conservation is enforced.

The main purpose of this paper is to show how the analytic control over the full probability distribution can be regained, and that in this way the reweighting approach becomes feasible. In the next section we review the basic parton shower and set up our notation. In Sec.~\ref{sec:standardPS} we take the parton shower of \sherpa~\cite{Kuhn:2000dk}, which has the same algorithm as \pythia's virtuality-ordered parton shower~\cite{Bengtsson:1986hr,Norrbin:2000uu,Sjostrand:2000wi}, as an example to investigate a real parton shower algorithm in detail. We then show in Sec.~\ref{sec:analyticPS} how it can be modified to satisfy four-momentum conservation at each vertex, while at the same time retaining the analytic control of the probability distribution. In Sec.~\ref{sec:reweighting} we discuss how these results can be applied in the different contexts described above, and Sec.~\ref{sec:conclusions} contains our conclusions.

%%%%%%%%%%%%%%%%%%%%%%%%%%%%%%%%%%%%%%%%%%%%%%%%%%%%%%%%%%%%%%%%%%%%%%%%%%%%%%%%
\section{Setup}
\label{sec:setup}
%%%%%%%%%%%%%%%%%%%%%%%%%%%%%%%%%%%%%%%%%%%%%%%%%%%%%%%%%%%%%%%%%%%%%%%%%%%%%%%%

%===============================================================================
\subsection{Branching Probabilities}
%===============================================================================

The purpose of this section is to review some of the basics of parton showers~\cite{Sjostrand:2006su} needed in our discussion and, in the process, introduce our notation. To be specific, we consider a final-state parton shower with the invariant mass as evolution variable. For simplicity, we assume all particles to be massless, although particle masses can be included in a straightforward way.

The branching of a mother parton with some energy $E$ into two daughter partons is determined by two independent kinematic variables, which are chosen to be the invariant mass of the mother $t$ (or equivalently the total invariant mass of the daughters) and the energy splitting $z$, which determines how the mother's energy is distributed between the daughters. Before the mother is branched, it is still on shell with $t = 0$. During the branching step the mother is put off shell and given an invariant mass $t > 0$. At the same time the energy splitting $z$ is obtained.

The single branch probability $\cP(t, z)$, defined as the differential probability for a branching to occur with certain values $t$ and $z$, is given by
%%%
\begin{align}
\label{Psingle}
\cP(t, z) &= f(t, z) \exp \biggl\{
-\int_t^{t_\mx} \!\df t' \!\int_{\frac{1}{2}-z_\cut}^{\frac{1}{2}+z_\cut} \!\df z'\, f(t', z') \biggr\}
\nn\\
&\equiv f(t, z)\, \Pi(t, t_\mx)
\,,\end{align}
%%%
where $f(t, z)$ is the usual Altarelli-Parisi splitting function~\cite{AP} and $\Pi(t, t_\mx)$ is the well-known Sudakov factor~\cite{Sudakov}, which resums the leading Sudakov double logarithms. The exact form of $t_\mx$ and $z_\cut$ in Eq.~\eqref{Psingle} depends on the details of the parton shower implementation and will be discussed later.

The algorithm to determine the value of $t$ at which the branching occurs is thought of as evolving $t$ from some high start value $t_\mx$ down to some lower value. The Sudakov factor $\Pi(t, t_\mx)$ then corresponds to the no-branching probability, i.e., the probability that no branching between $t_\mx$ and $t$ occurs. To see this, one integrates $\cP(t,z)$,
%%%
\begin{equation}
\int_{t_1}^{t_\mx} \!\df t \int_{\frac{1}{2}-z_\cut}^{\frac{1}{2}+z_\cut} \!\df z\, \cP(t, z)
 = 1 - \Pi(t_1, t_\mx)
\,.\end{equation}
%%%
The left-hand side is the probability for the mother to branch somewhere between $t_1$ and $t_\mx$. Since the total probability is unity, $\Pi(t_1, t_\mx)$ is the probability that the mother does not branch between $t_1$ and $t_\mx$.

When the mother is branched, the daughters are still on shell with zero invariant mass. In the next step the daughters themselves are branched and given non zero invariant masses, and after that, their daughters, and so on, resulting in a treelike structure as shown in Fig.~\ref{fig:tree}. As the invariant mass of each daughter must be less than that of its mother, $t$ is always decreasing for consecutive branchings. When a branching with $t < t_\cut$ is obtained, the corresponding mother is left unbranched. The parton shower terminates once no more branchings with $t \geq t_\cut$ are found. The value of the cutoff $t_\cut$ is typically chosen to be a low scale of order a few $\GeV^2$.

Each event produced by the parton shower consists of a tree of $n$ branches characterized by the set of variables $\{t_i, z_i\} \equiv \{t_1, z_1; t_2, z_2; \ldots; t_n, z_n \}$, which can later be turned into momentum four-vectors. As discussed in the Introduction, we would like to be able to explicitly compute the probability for a given event with certain values $\{t_i, z_i\}$ to be generated. Ideally, this probability is simply given as the product of individual single branch probabilities, schematically,
%%%
\begin{equation}
\label{prob_product}
P_\PS(\{t_i, z_i\})
= \cP(t_1, z_1) \times \cP(t_2, z_2) \times \dotsb
\,.\end{equation}
%%%
If the parton shower satisfies Eq.~\eqref{prob_product}, the problem of finding a closed form for $P_\PS(\{t_i, z_i\})$ reduces to working out the exact form of the single branch probability $\cP(t, z)$.

It is important to point out that, even though the parton shower itself is only valid in the limit where the invariant mass of each daughter is much smaller than that of its mother, we still need to know the exact form of $P_\PS(\{t_i, z_i\})$ for all values of $\{t_i,z_i\}$. In other words, it is not sufficient for our purposes to only know $P_\PS(\{t_i, z_i\})$ expanded in the limit where the parton shower is valid.

Eq.~\eqref{prob_product} would be trivially satisfied if all branchings were independent of each other. In general, this is not the case for two reasons. First, the branching of each daughter depends on the initial conditions set by its mother, which implies that $P_\PS(\{t_i, z_i\})$ depends on the specific structure of the tree. However, as long as each branching only depends on previous branchings, the total probability can still be written as a product of single branch probabilities as in Eq.~\eqref{prob_product}.

The second reason Eq.~\eqref{prob_product} can be violated is more involved and related to the basic issue of the parton shower we wish to address. At the time each branch is generated, the corresponding daughters have not yet been branched and are still on shell. However, the phase-space limits for the branch following from four-momentum conservation depend on the daughters' final invariant masses. Hence, only after both daughters have been branched themselves, one can come back and enforce the correct phase-space limits on the branch. Usually this step involves some kinematic reshuffling, which ends up introducing a complicated correlation between the daughters' branches and thereby violating Eq.~\eqref{prob_product}.

%===============================================================================
\subsection{Kinematics}
%===============================================================================

\begin{figure}[t]
\centering
\includegraphics[width=0.65\columnwidth]{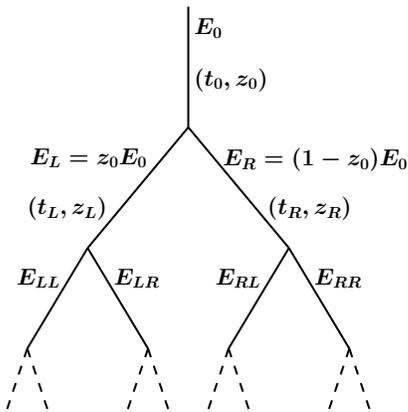}
\caption{Diagrammatic representation of a tree of branches.}
\label{fig:tree}
\end{figure}

We begin by working out the phase space for a single $1\to 2$ branch. We denote all kinematic quantities (see Fig.~\ref{fig:tree}) related to the mother particle with a subscript $\mo$ and those of the left and right daughters with a subscript $L$ and $R$, respectively. We regard the mother's invariant mass $t_\mo$ and energy $E_\mo$ as fixed and take the daughters to have invariant masses $t_L$ and $t_R$. The energy splitting $z_\mo$ is regarded as a property of the mother (or rather of the whole branch) linking the daughters' energies to $E_\mo$,
%%%
\begin{equation}
\label{zmo_def}
E_L = z_\mo E_\mo
\,,\qquad
E_R = (1 - z_\mo) E_\mo
\,.\end{equation}
%%%
Thus, energy conservation is automatically satisfied, and $E_L$ and $E_R$ are not free variables.

In the rest frame of the mother, the value of $z_\mo$ is fixed in terms of $t_L$ and $t_R$,
%%%
\begin{equation}
\label{zmo_CM}
z_\mo^\mathrm{CM} = \frac{1}{2} \biggl( 1 + \frac{t_L}{t_\mo} - \frac{t_R}{t_\mo} \biggr)
\,.\end{equation}
%%%
Using Eqs.~\eqref{zmo_def} and \eqref{zmo_CM}, the magnitude of the daughters' three-momenta is
%%%
\begin{equation}
\label{pvec}
\mathbf{p}_\mathrm{CM}^2 = E_L^2 - t_L = E_R^2 - t_R = \frac{t_\mo}{4} \lambda(t_\mo, t_L, t_R)^2
\,,\end{equation}
%%%
with the usual phase-space factor
%%%
\begin{equation}
\label{lambda}
\lambda(t_\mo, t_L, t_R) = \frac{1}{t_\mo} \sqrt{(t_\mo - t_L - t_R)^2 - 4t_L t_R}
\,.\end{equation}
%%%
To enforce timelike daughters one needs
%%%
\begin{equation}
t_L \leq E_L^2
\,,\qquad
t_R \leq E_R^2
\,,\end{equation}
%%%
which, using Eqs.~\eqref{pvec} and~\eqref{lambda}, is equivalent to the usual phase-space limit
%%%
\begin{equation}
\sqrt{t_L} + \sqrt{t_R} \leq \sqrt{t_\mo}
\,.\end{equation}
%%%

The kinematics in a general frame is obtained by boosting the above results. The magnitude of the boost $\beta_\mo = \sqrt{1 - t_\mo/E_\mo^2}$ is fixed by $E_\mo$ and $t_\mo$, while its direction can be described by the angle $\theta_\mo$ between the boost axis and the daughters' three-momenta in the mother's rest frame. Since $\theta_0$ encodes the information how the energies of the daughters are boosted relative to the mother's rest frame, it effectively determines $z_\mo$,
%%%
\begin{align}
\label{zmo_GF}
z_\mo &= z_\mo^\mathrm{CM} + \beta_\mo \cos\theta_\mo\, \frac{\lvert\mathbf{p}_\mathrm{CM}\rvert}{\sqrt{t_\mo}}
\nn\\
&= \frac{1}{2} \biggl[1 + \frac{t_L}{t_\mo} - \frac{t_R}{t_\mo}
+ \beta_\mo \cos\theta_\mo \lambda(t_\mo, t_L, t_R) \biggr]
,\end{align}
%%%
and vice versa,
%%%
\begin{equation}
\label{costhetamo}
\cos\theta_\mo
= \frac{1}{\beta_\mo}\, \frac{2z_\mo - (1 + t_L/t_\mo - t_R/t_\mo)}{\lambda(t_\mo, t_L, t_R)}
\,.\end{equation}
%%%
The phase-space limits in a general frame are a simple generalization of the limits in the rest frame,
%%%
\begin{equation}
\label{phsp_single}
\sqrt{t_L} + \sqrt{t_R} \leq \sqrt{t_\mo}
\,,\qquad
\lvert\cos\theta_\mo\rvert \leq 1
\,.\end{equation}
%%%
Using Eq.~\eqref{costhetamo}, the limit on $\cos\theta_\mo$ is equivalent to the $z_\mo$ limit
%%%
\begin{equation}
\label{zmo_limits}
\biggl\lvert z_\mo - \frac{1}{2}\biggl(1 + \frac{t_L}{t_\mo} - \frac{t_R}{t_\mo} \biggr) \biggr\rvert
\leq \frac{\beta_\mo}{2} \lambda(t_\mo, t_L, t_R)
\,,\end{equation}
%%%
commonly found in parton shower algorithms.

Finally, we look at a double branch $1\to 2 \to 4$ where each daughter further branches into two on-shell particles. In this case, the daughters' energy splittings $z_{L,R}$, or equivalently $\theta_{L,R}$, are additional free variables. The complete phase space now is just an extension of Eq.~\eqref{phsp_single},
%%%
\begin{equation}
\label{phsp_double}
\sqrt{t_L} + \sqrt{t_R} \leq \sqrt{t_\mo}
\,,\quad
\lvert\cos\theta_\mo\rvert \leq 1
\,,\quad
\lvert\cos\theta_{L,R} \rvert \leq 1
\,.\end{equation}
%%%
The limits on $z_{L,R}$ equivalent to $\lvert\cos\theta_{L,R}\rvert \leq 1$ are analogous to Eq.~\eqref{zmo_limits} with the daughters' invariant masses set to zero. Hence, the complete phase space in terms of $t_{L,R}$ and $z_{0,L,R}$ reads
%%%
\begin{align}
\label{phsp_doublez}
\sqrt{t_L} + \sqrt{t_R} &\leq \sqrt{t_\mo}
\,,\nn\\
\biggl\lvert z_\mo - \frac{1}{2}\biggl(1 + \frac{t_L}{t_\mo} - \frac{t_R}{t_\mo} \biggr) \biggr\rvert
&\leq \frac{\beta_\mo}{2} \lambda(t_\mo, t_L, t_R)
\,,\nn\\
\biggl\lvert z_L - \frac{1}{2} \biggr\rvert
&\leq \frac{\beta_L}{2}
\,,\nn\\
\biggl\lvert z_R - \frac{1}{2} \biggr\rvert
&\leq \frac{\beta_R}{2}
\,,\end{align}
%%%
with
\begin{equation}
\label{betaLR_def}
\beta_i = \sqrt{1 - t_i/E_i^2}
\,,\end{equation}
and $E_{L,R}$ as in Eq.~\eqref{zmo_def}.

Eq.~\eqref{phsp_doublez} explicitly shows the problem mentioned at the end of the previous section. Initially, $z_\mo$ is generated assuming $t_{L,R} = 0$, but since the limit on $z_\mo$ depends on $t_L$ and $t_R$, the generated value of $z_\mo$ has to be adjusted after $(t_L,z_L)$ and $(t_R,z_R)$ have been determined. Changing $z_\mo$, however, changes $E_{L,R}$ and $\beta_{L,R}$. This in turn changes the limits on $z_L$ and $z_R$, which can render their values invalid. In addition, $t_L$ and $t_R$ are determined independently from one another, so the constraint $\sqrt{t_L} + \sqrt{t_R} \leq \sqrt{t_\mo}$ can be violated as well.

%%%%%%%%%%%%%%%%%%%%%%%%%%%%%%%%%%%%%%%%%%%%%%%%%%%%%%%%%%%%%%%%%%%%%%%%%%%%%%%%
\section{A Standard Parton Shower}
\label{sec:standardPS}
%%%%%%%%%%%%%%%%%%%%%%%%%%%%%%%%%%%%%%%%%%%%%%%%%%%%%%%%%%%%%%%%%%%%%%%%%%%%%%%%

To study a concrete example of a standard parton shower, we consider the final-state parton shower of \sherpa~\cite{Kuhn:2000dk}, which employs the same algorithm as the \pythia\ virtuality-ordered parton shower~\cite{Bengtsson:1986hr,Norrbin:2000uu,Sjostrand:2000wi}. Other algorithms which employ different ordering variables can be found in Refs.~\cite{Marchesini:1983bm,Gustafson:1987rq,Gieseke:2003rz,Sjostrand:2004ef}.

%===============================================================================
\subsection{The Algorithm}
%===============================================================================

\begin{figure*}[t!]
\centerline{\includegraphics[width=0.75\textwidth]{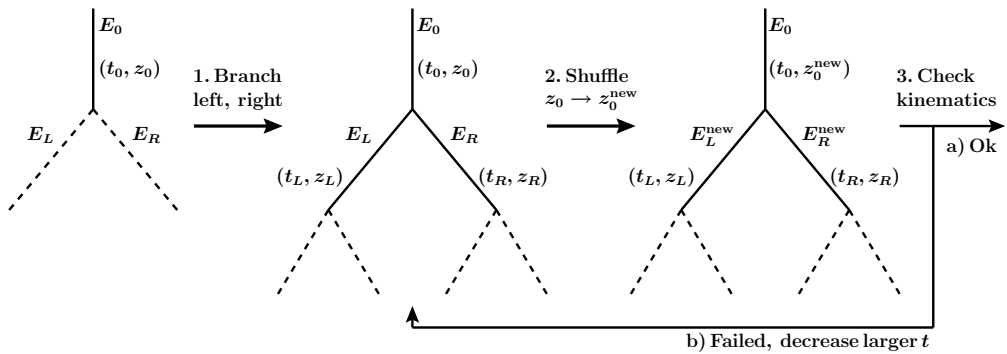}}
\caption{Diagrammatic representation of a standard parton shower algorithm. Solid lines represent off-shell partons with nonzero invariant mass, dashed lines unbranched, on-shell partons.}
\label{fig:alg1}
\end{figure*}

To be able to enforce four-momentum conservation, the parton shower algorithm always branches two sisters in pairs. That is, in each iteration it takes an existing $1\to 2$ branch, consisting of a mother and two unbranched daughters, and converts it into a $1\to2\to4$ double branch by branching both daughters. To do so, the algorithm proceeds in three steps as depicted in Fig.~\ref{fig:alg1}:
%%%
\begin{enumerate}
\item Branch both daughters, each according to $\cP(t, z)$.
\item Shuffle $z_\mo \to z_\mo^\new(z_\mo, t_L, t_R)$.
\item Check kinematics in terms of new $z_\mo^\new$:
  \begin{enumerate}
  \item If successful, accept daughter branches.
  \item If failed, evolve daughter with larger $t$ further down and return to step 2.
  \end{enumerate}
\end{enumerate}
%%%
In step 1, each daughter is branched separately, with values for $(t_L, z_L)$ and $(t_R, z_R)$ distributed according to the single branch probability $\cP(t, z)$. In step 2, $z_\mo$ is changed to a new value $z_\mo^\new(z_\mo, t_L, t_R)$, which is derived from its old value and takes into account the now nonzero values of $t_{L,R}$.
In the mother's restframe this shuffling simply sets $z_\mo$ to the correct value $z_\mo^\mathrm{CM}$ in Eq.~\eqref{zmo_CM}. In a general frame the form of $z_\mo^\new(z_\mo, t_L, t_R)$ is not dictated by kinematics anymore, but is usually chosen to satisfy Eq.~\eqref{zmo_limits}. In step 3, the kinematics are checked, using the new value $z_\mo^\new$. If they are satisfied, the daughter branches are accepted. Otherwise, the algorithm takes the daughter with the larger $t$, evolves it further down, and goes back to step 2.

In the remainder of this section we discuss the details of this algorithm. We first work out the precise form of the single branch probability $\cP(t, z)$ employed by the algorithm. We then move to discuss in detail steps 2 and 3, which implement four-momentum conservation, but as one can already see from Fig.~\ref{fig:alg1} and the above discussion, introduce a complicated correlation between $t_L$ and $t_R$, which clearly violates Eq.~\eqref{prob_product}.

%===============================================================================
\subsection{The Single Branch Probability}
%===============================================================================

In this section we are only interested in a single $1\to2$ branching of a mother into two daughters. This means that each of the daughters in the algorithm described above acts as the mother now, and similarly, the mother and the other daughter in the algorithm now act as grandmother and sister, respectively.

In the first step, the algorithm independently generates two sets of values $(t_L,z_L)$ and $(t_R,z_R)$ according to the single branch probability $\cP(t, z)$, introduced in Eq.~\eqref{Psingle}. The precise form of $\cP(t, z)$ that is actually used in the algorithm can be written as
%%%
\begin{align}
\label{Psingle_precise}
\cP(t, z) &= f(t, z)\, \Pi(t, t_\mx)\, \theta(t_\mx - t)\, \theta(t - t_\cut)
\nn\\
& \quad \times
\theta\Bigl[z_\cut(t) - \Bigl\lvert z - \frac{1}{2}\Bigr\rvert\Bigr]
\,,\end{align}
%%%
with the Sudakov factor
%%%
\begin{equation}
\label{Sud_precise}
\Pi(t_1, t_2) = \exp \biggl\{-\int_{t_1}^{t_2} \!\df t \int_{\frac{1}{2} - z_\cut(t)}^{\frac{1}{2} + z_\cut(t)}\!\df z\, f(t, z) \biggr\}
\,.\end{equation}
%%%
In Eq.~\eqref{Psingle_precise} we explicitly included all kinematic $\theta$ functions restricting the allowed ranges of $t$ and $z$. All information on the precise form of $\cP(t,z)$ is encoded in the integration limits $t_\mx$ and $z_\cut(t)$, and the splitting function $f(t,z)$. Note that $\cP(t, z)$ is not normalized to unity, because it describes the differential distribution in $(t, z)$ for an entire $1\to 2$ branch. It does not include the probability that the mother parton does not branch, in which case $z$ is undefined. However, we can still define the differential distribution in $t$ for a single parton,
%%%
\begin{equation}
\label{Psingle_normalized}
\cP(t) = \Pi(t_\cut, t_\mx)\, \delta(t) + \int\!\df z\, \cP(t, z)
\,,\end{equation}
%%%
where the first term is the no-branching probability, and $\cP(t)$ is now properly normalized to unity,
%%%
\begin{equation}
\int\!\df t\, \cP(t) = \Pi(t_\cut, t_\mx) + [1 - \Pi(t_\cut, t_\mx)] = 1
\,.\end{equation}
%%%

The precise form of the splitting function $f(t,z)$ depends on the specific type of splitting ($q\to qg$, $g\to qq$, or $g\to gg$), e.g.
%%%
\begin{equation}
\label{fqqg}
f_{q\to qg}(t, z) = \frac{\alpha_s(\mu) C_F}{2\pi}\, \frac{1}{t}\, \frac{1 + z^2}{1 - z}
\,.\end{equation}
%%%
The scale at which $\alpha_s$ is evaluated is, in general, a function of $t$ and $z$. For simplicity we use $\mu^2 = t/4$. Another typical choice is $\mu^2 = z(1-z)t$.

We stress that the kinematics relevant to Eq.~\eqref{Psingle_precise} assumes that both daughters and the mother's sister are on-shell, massless particles. The expressions for $t_\mx$ and $z_\cut(t)$ arising from the phase space limits are then
%%%
\begin{equation}
\label{tmaxzcut_phsp}
t_\mx = E_\ini^2
\,,\qquad
z_\cut(t) = \frac{\beta}{2}
\,,\end{equation}
%%%
where $E_\ini$ is the initial energy of the mother%
\footnote{Usually, $E_\ini$ is given in terms of the grandmother's $z_\mo$ and $E_\mo$. One exception is the case when the mother is the final parton coming from a hard interaction in the grandmother's rest frame. In this case, $E_\ini$ is chosen to be $E_\ini(t) = \sqrt{t_\mo}(1 + t/t_\mo)/2$, which is the exact result for an on-shell, massless sister.}%
, and $\beta = \sqrt{1 - t/E_\ini^2}$ in this subsection.

In addition to the pure phase space limits, the algorithm includes several restrictions which modify the form of $t_\mx$ and $z_\cut(t)$. First, since the parton showers is ordered in the evolution variable, $t$ is always smaller than the initial value $t_\ini$ where the evolution starts, which is usually chosen to be the invariant mass of the grandmother, $t_\ini = t_\mo$. Second, the cutoff on the algorithm, $t \geq t_\cut$, is realized as a cut on $p_T^2 \equiv \lvert\mathbf{p}_T\rvert^2 \geq t_\cut/4$, where $\mathbf{p}_T$ is the daughters' transverse momentum with respect to the mother's flight direction. In terms of $t$ and $z$ we have
%%%
\begin{equation}
p_T^2 = \frac{t}{4} - \frac{t}{\beta^2}\Bigl(z - \frac{1}{2}\Bigr)^2
\,.\end{equation}
%%%
Hence, $p_T^2 \geq t_\cut/4$ implies $t \geq t_\cut$, but with the additional restriction
%%%
\begin{equation}
\biggl\lvert z - \frac{1}{2} \biggr\rvert
\leq \frac{\beta}{2} \sqrt{1 - \frac{t_\cut}{t}}
\,.\end{equation}
%%%

Furthermore, parton showers require the opening angle between the daughters of subsequent emissions to always decrease. This angular ordering ensures that the branching is correctly described not only in the collinear limit, where the splitting functions are derived, but also in the soft limit, where the branching is coherent~\cite{Mueller:1981ex,Bengtsson:1986hr}. The opening angle $\vartheta$ is given by
%%%
\begin{equation}
\cos\vartheta = 1 - \frac{t}{2z(1-z) E_\ini^2}
\,,\end{equation}
%%%
which allows us to translate the angular ordering cut $\vartheta \leq \vartheta_\cut$, where $\vartheta_\cut$ is the opening angle of the previous branch, into limits on $t$ and $z$,
%%%
\begin{align}
t &\leq E_\ini^2 \frac{1-\cos\vartheta_\cut}{2}
\,,\nn\\
\biggl\lvert z - \frac{1}{2} \biggr\rvert
&\leq \frac{\beta}{2} \sqrt{1 - \frac{t}{\beta^2 E_\ini^2}\,\frac{1+\cos\vartheta_\cut}{1-\cos\vartheta_\cut}}
\,.\end{align}
%%%

Putting everything together, the single branch probability $\cP(t,z)$ depends on several additional parameters restricting the allowed range of $t$ and $z$,
%%%
\begin{equation}
\cP(t,z) \equiv \cP(t,z\vert t_\ini,E_\ini,t_\cut,\vartheta_\cut)
\,,\end{equation}
%%%
where the limits on $t$ and $z$ are determined by
%%%
\begin{align}
\label{tmaxzcut_full}
t_{\rm max} &= \min\biggl\{t_\ini\,,E_\ini^2\frac{2}{1-\cos\vartheta_\cut}\biggr\}
\,,\\\nn
z_\cut(t) &= \frac{\beta}{2} \min\biggl\{\sqrt{1 - \frac{t}{\beta^2 E^2_i}\,\frac{1+\cos\vartheta_\cut}{1-\cos\vartheta_\cut}}\,,\sqrt{1 - \frac{t_\cut}{t}}\biggr\}
\,.\end{align}
%%%
Note that these limits automatically include the phase-space limits, Eq.~\eqref{tmaxzcut_phsp}, which are reproduced for $\vartheta_\cut = \pi$ and $t_\cut = 0$. Also, for $t \to E^2$, corresponding to the mother's rest frame, $z_\cut(t) \to 0$, forcing $z \to 1/2$, as required.

%===============================================================================
\subsection{Enforcing Momentum Conservation}
%===============================================================================

In each iteration the algorithm in Fig.~\ref{fig:alg1} starts with a value for $z_\mo$ which, as discussed above, was determined in a previous iteration assuming $t_{L,R} = 0$ and satisfying $\lvert 2z_\mo -1 \rvert \leq \beta_\mo$. In particular, in the mother's rest frame it starts with $z_\mo = 1/2$, whereas after step 1, $t_L$ and $t_R$ have become nonzero and so the correct value is now $z_\mo^{\mathrm CM}$ given in Eq.~\eqref{zmo_CM}. Thus, the value of $z_\mo$ has to be changed (shuffled), or otherwise the kinematics can never be satisfied.

This shuffling happens in step 2 of the algorithm. There are various ways to do this, and \sherpa\ uses
%%%
\begin{align}
\label{znew}
z_\mo &\to z_\mo^\new(z_\mo,t_L,t_R)
\nn\\ & \quad
= \frac{1}{2} \biggl[ 1 + \frac{t_L}{t_\mo} - \frac{t_R}{t_\mo} + (2 z_\mo - 1) \lambda(t_\mo,t_L,t_R) \biggr]
\,,\end{align}
%%%
with $\lambda(t_\mo,t_L,t_R)$ given in Eq.~\eqref{lambda}. For $z_\mo = 1/2$, Eq.~\eqref{znew} reduces to Eq.~\eqref{zmo_CM}, as required. Since the original value of $z_\mo$ satisfies $\lvert 2z_\mo - 1 \rvert \leq \beta_\mo$, the shuffling in Eq.~\eqref{znew} ensures that $z_\mo$ always satisfies the correct phase-space limit, Eq.~\eqref{zmo_limits}, for $t_{L,R} \neq 0$. Nevertheless, four-momentum conservation can still be violated in two ways. First, it may not be possible at all to find a new physical value $z_\mo^\new$. Namely, $t_{L,R}$ may not satisfy the constraint $\sqrt{t_L} + \sqrt{t_R} \leq \sqrt{t_0}$ in Eq.~\eqref{phsp_doublez}, which ensures timelike daughters and that $\lambda(t_\mo, t_L, t_R)$ in Eq.~\eqref{znew} is well defined. This can happen, because the values of $t_L$ and $t_R$ were determined independently from one another, only subject to the constraint $t_{L,R} \leq t_\mx$ [see Eq.~\eqref{tmaxzcut_full}].

Second, changing $z_\mo \to z_\mo^\new$ also changes the energies of the two daughters,
%%%
\begin{equation}
\label{ELRnew}
E_L \to E_L^\new = z_\mo^\new E_\mo
\,,\quad
E_R \to E_R^\new = (1 - z_\mo^\new) E_\mo
\,,\end{equation}
%%%
and accordingly [see Eq.~\eqref{betaLR_def}],
%%%
\begin{equation}
\label{betaLRnew}
\beta_{L,R} \to \beta_{L,R}^\new = \sqrt{1 - t_{L,R}/E_{L,R}^\new}
\,.\end{equation}
%%%
Now assume, for instance, that $t_L > t_R$. Then $z_\mo^{\rm new} > z_\mo$, which increases $E_L$ (decreases $E_R$) and decreases $\beta_L$ (increases $\beta_R$). Using Eq.~\eqref{phsp_doublez}, it follows that the available phases space for $z_L$ shrinks, and hence the value of $z_L$ may not be allowed anymore. The same is true for the right branch in case $t_R > t_L$.

For this reason, the algorithm has to explicitly check the kinematics, which is done in step 3. In \sherpa\ this is implemented by checking various different kinematical constraints arising from four-momentum conservation, all of which can be reduced to the constraints in Eq.~\eqref{phsp_doublez}. If all constraints are satisfied, the daughter branches are accepted and the algorithm proceeds with the next iteration. Otherwise, if any phase-space limit is violated, the algorithm picks the daughter with the larger $t$, generates new values $(t, z)$ according to $\cP(t, z)$ with the previous $t$ as $t_\ini$, and goes back to step 2.

It should be clear from this discussion that steps 2 and 3 in the algorithm introduce a complicated cross correlation between the probabilities to get certain values $(t_L,z_L)$ and $(t_R,z_R)$, which breaks the factorization in Eq.~\eqref{prob_product}, and makes it virtually impossible to find a closed-form expression for the double branch probability $P(t_L,z_L; t_R,z_R)$.

To end this section, we note that the shuffling of $z_\mo$ is proportional to $t_{L,R}/t_\mo$. As the parton shower is formally only valid for $t_{L,R} \ll t_\mo$, this is formally a power suppressed effect. Four-momentum conservation, however, is an important power correction that must be taken into account in order to obtain realistic events, also because in the end the parton shower is used for any values $t_{L,R} \leq t_\mo$.

%%%%%%%%%%%%%%%%%%%%%%%%%%%%%%%%%%%%%%%%%%%%%%%%%%%%%%%%%%%%%%%%%%%%%%%%%%%%%%%%
\section{The Analytic Parton Shower}
\label{sec:analyticPS}
%%%%%%%%%%%%%%%%%%%%%%%%%%%%%%%%%%%%%%%%%%%%%%%%%%%%%%%%%%%%%%%%%%%%%%%%%%%%%%%%

%===============================================================================
\subsection{The Analytic Algorithm}
%===============================================================================

\begin{figure*}[tp]
\centerline{\includegraphics[width=0.75\textwidth]{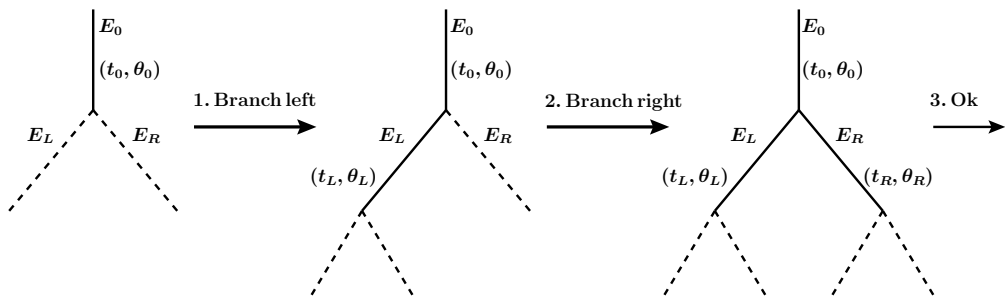}}
\caption{Diagrammatic representation of the analytic algorithm. Solid lines represent off-shell partons with nonzero invariant mass, dashed lines unbranched, on-shell partons. The evolution of $t_L$ in step 1 starts at $t_\ini = t_\mo$, and that of $t_R$ in step 2 at $t_\ini = (\sqrt{t_\mo} - \sqrt{t_L})^2$.}
\label{fig:alg2}
\end{figure*}

It is the final step in the parton shower, where a generated set $(t_L,z_L)$ or $(t_R,z_R)$ can be rejected, which leads to the complicated correlation in the double branch probability. As explained above, there are two reasons the kinematics can fail in step 3. First, it is possible that the original values of $t_{L,R}$ violate $\sqrt{t_L} + \sqrt{t_R} \leq \sqrt{t_0}$, and second, the required momentum shuffling in step 2 can render the values of $z_{L,R}$ invalid. We now show how both of these problems can be dealt with without introducing complicated correlations.

We start by looking at the second problem, for which it is instructive to understand in more detail the physical picture behind the shuffling in Eq.~\eqref{znew}. Looking back at Eq.~\eqref{zmo_GF}, one can think of $z_\mo$ in a general frame as given in terms of $t_{\mo,L,R}$ and $\beta_\mo \cos\theta_\mo$. Vice versa, for given $t_{\mo,L,R}$, $z_\mo$ determines the boost factor $\beta_\mo \cos\theta_\mo$. When $z_\mo$ was generated, $t_{L,R} = 0$, and Eq.~\eqref{zmo_GF} implies
%%%
\begin{equation}
\beta_\mo \cos\theta_\mo = 2z_\mo - 1
\,.\end{equation}
%%%
Thus, selecting a value for $z_\mo$ is equivalent to choosing the boost factor $\beta_\mo\cos\theta_\mo$. With this in mind, it is easy to understand what the shuffling in Eq.~\eqref{znew} is doing physically. It simply holds $\beta_\mo\cos\theta_\mo$ fixed at its generated value, and recomputes $z_\mo$ using Eq.~\eqref{zmo_GF} after the daughters acquire nonzero invariant masses. In other words, the algorithm treats the boost factor $\beta_\mo\cos\theta_\mo$ as the fundamental and $z_\mo$ as the derived quantity.%
\footnote{This statement is true in general, even if $z_\mo$ is shuffled more than once, because in the algorithm the value of $z_\mo$ used on the right-hand side of Eq.~\eqref{znew} is always the originally generated value, never one which was already obtained from shuffling.}

In the same way, generating values for $z_{L,R}$ in step 1 in Fig.~\ref{fig:alg1} can be regarded as generating and fixing values for the boost factors
%%%
\begin{equation}
\label{betacosthetaLR}
\beta_{L,R} \cos\theta_{L,R} = 2 z_{L,R} - 1
\,.\end{equation}
%%%
However, $\beta_{L,R}$ are not free quantities, but change with $z_\mo$ as functions of $t_{L,R}$ [see Eqs.~\eqref{ELRnew} and \eqref{betaLRnew}]. The only way to keep $\beta_{L,R}\cos\theta_{L,R}$ fixed when $z_\mo$ is shuffled is to balance the change in $\beta_{L,R}$ by a corresponding change in $\cos\theta_{L,R}$. In this picture, we now immediately see what can go wrong. When shuffling $z_\mo$, either $\beta_L$ or $\beta_R$ might decrease too much, resulting in an unphysical value $\lvert\cos\theta\rvert > 1$ for either $\theta_L$ or $\theta_R$.

This picture also leads to a simple and general solution to the problem: Instead of $\beta\cos\theta$, we can just as well use the boost angle $\theta$ itself as the fundamental quantity. When a value for $z$ is generated, it is translated into a value for $\cos\theta$,
%%%
\begin{equation}
\label{costheta}
\cos\theta = \frac{2 z - 1}{\beta}
\,,\end{equation}
%%%
which is held fixed at any later stage in the algorithm. The advantage is that the phase-space limits $\lvert\cos\theta\rvert \leq 1$ are completely independent of any other kinematic variables and are always satisfied. Holding $\cos\theta$ fixed in Eq.~\eqref{costheta} corresponds to an additional shuffle
%%%
\begin{equation}
\label{znew_LR}
z \to z^\new(z,\beta,\beta^\new) = \frac{1}{2} \biggl[1 + (2z - 1) \frac{\beta^\new}{\beta} \biggr]
\,,\end{equation}
%%%
to be applied to $z_{L,R}$ whenever $\beta_{L,R}$ change as a result of shuffling $z_\mo$. Similarly to Eq.~\eqref{znew}, Eq.~\eqref{znew_LR} ensures that $z_{L,R}$ always satisfy their phase-space constraints.

Note that Eqs.~\eqref{znew} and \eqref{znew_LR} can be realized as two special cases of a generalized shuffle, which follows from Eqs.~\eqref{costheta} and \eqref{zmo_GF},
%%%
\begin{align}
\label{znew_general}
z^\old &\to z^\new(z^\old,\beta^\old, t_L,t_R,\beta^\new)
\\\nn & \quad
= \frac{1}{2} \biggl[ 1 + \frac{t_L}{t} - \frac{t_R}{t} + (2 z^\old - 1) \frac{\beta^\new}{\beta^\old} \lambda(t,t_L,t_R) \biggr]
\,,\end{align}
%%%
where $\beta^\old$ is the value used when $z^\old$ was generated, and $\beta^\new$ is the new value. This makes the practical implementation into the current algorithms straightforward, because all one has to do is to store $\beta^\old$ and change the shuffling function.

Since all Eq.~\eqref{znew_general} does is to hold $\cos\theta$ fixed, we can also go one step further and directly use $(t,\theta)$ to describe the individual branches, with the daughters' energies $E_{L,R}$ always given as functions of $E_\mo$, $\theta_\mo$ and $t_{\mo,L,R}$. In this way, any necessity to keep track of $z_{\mo,L,R}$ and how and when they are shuffled is removed, which makes the entire algorithm very transparent.

At this point, the only kinematical check that would be left in step 3 is the simple constraint $\sqrt{t_L} + \sqrt{t_R} \leq \sqrt{t_\mo}$, which cannot be eliminated by a change of variables. However, we can recast the correlation it introduces into a calculable form by branching the daughters in two separate steps, as shown in Fig.~\ref{fig:alg2}. In the first step the left daughter is branched, starting the evolution at $t_\ini = t_\mo$. In the second step the right daughter is branched, starting the evolution at $t_\ini = (\sqrt{t_\mo} - \sqrt{t_L})^2$, which automatically takes care of the remaining constraint. In this way, the double branch probability for the left and right branches can be written as the product of two single branch probabilities
%%%
\begin{align}
\label{Pdouble}
P(t_L,\theta_L; t_R, \theta_R) &= \cP(t_L, \theta_L \vert t_\ini = t_\mo)
\\\nn
&\quad \times \cP\bigl[t_R, \theta_R \vert t_\ini = (\sqrt{t_\mo} - \sqrt{t_L})^2\bigr]
\,.\end{align}
%%%
Since the iteration steps in the generation of an entire event are already independent, Eq.~\eqref{Pdouble} allows us to write the total probability to generate a given event as a product of single branch probabilities as in Eq.~\eqref{prob_product}, which is what we set out to do.

%===============================================================================
\subsection{Practical Implementation}
%===============================================================================

In practice, one has several choices in implementing this new analytic algorithm in a real parton shower. To eliminate the asymmetry between $t_L$ and $t_R$ in Eq.~\eqref{Pdouble}, one can randomly choose which daughter of a given branch acts as the left daughter and is branched first. Alternatively, one can always branch the daughter with the larger (or smaller) initial energy first. A third choice is to generate a test value of $t$ for each daughter, call the larger one $t_L$, and then branch the other daughter starting the evolution at the smaller of $t_L$ and $(\sqrt{t_\mo}-\sqrt{t_L})^2$. For this choice, the double branch probability still factorizes and involves an additional factor of the no-branching probability $\Pi(t_\mo,t_L)$. This last choice may be the most natural one from the point of view of a global evolution~\cite{peter}.

Once the left daughter has been branched, one also has the choice which energy to use as the initial energy $E_\ini$ for branching the right daughter. One could either keep the original energy computed from $\theta_\mo$ with $t_L = 0$ or recompute it from $\theta_\mo$ with the new value of $t_L$, where again the latter choice seems to be the more natural one.

We have implemented the changes to the algorithm described above in \sherpa's parton shower. For simplicity, we always branch the left daughter first and keep the original energy for $E_\ini$ when branching the right daughter. The only things we had to change then were to separate the branching of the two daughters and to change the function returning a new value of $z$ to use Eq.~\eqref{znew_general}. As a cross check, we did not remove the kinematic checks done in step 3 of the original algorithm and tested that with our modifications they indeed never fail (apart from extremely rare occasions where the failure is due to numerical inaccuracies).

%===============================================================================
\subsection{Comparison and Numerical Results}
\label{subsec:comparison}
%===============================================================================

We now discuss the impact the change in the algorithm has on the generated events. The original algorithm rejects branches in the third step if the kinematics is not satisfied, which leads to a lowering of at least one of the invariant masses of the daughters that are branched. Hence, compared to the analytic algorithm, where this third step is absent, the original algorithm suppresses large values of $t$ and enhances low values of $t$. Since the effect on the kinematics from having nonzero $t$ is more pronounced for larger $t$, this relative suppression is expected to increase with increasing $t$.

To estimate the expected size of this effect, we note that shuffling $z$ is a power suppressed effect, of order $t/t_\mo$, which one can think of as changing the $z$ limits of integration. Thus, upon integration over $z$, the difference in the two algorithms corresponds to a power correction to the splitting function $f(t) \equiv \int\!\df z\, f(t,z)$, schematically
%%%
\begin{equation}
\label{Deltaf}
\Delta f(t) =  f(t) \times \ord\Bigl(\frac{t}{t_\mo}\Bigr)
\,.\end{equation}
%%%
Furthermore, the single branch probability generated by the algorithm always has the form $P(t) = f(t)\,\exp[{\int \df t\,f(t)}]$. The integral of a difference in $f(t)$ in the exponent gives rise to an additional finite perturbative difference. Thus, Eq.~\eqref{Deltaf} translates into a change in the single branch probability
%%%
\begin{equation}
\label{DeltaP}
\Delta P(t) =  P(t) \times \biggl[ \ord\Bigl(\frac{t}{t_\mo}\Bigr) + \ord(\alpha_s) \biggr]
\,.\end{equation}
%%%
The appearance of perturbative corrections can also be understood in another way. Since the total probability $P(t)$ is normalized to unity [see Eq.~\eqref{Psingle_normalized}], and power corrections must vanish for $t\to 0$, an increase of $P(t)$ at large $t$ via power corrections can only be compensated at small $t$ by a decrease of $P(t)$ via perturbative corrections.

\begin{figure*}[t]
\centerline{\includegraphics[width=.45\textwidth]{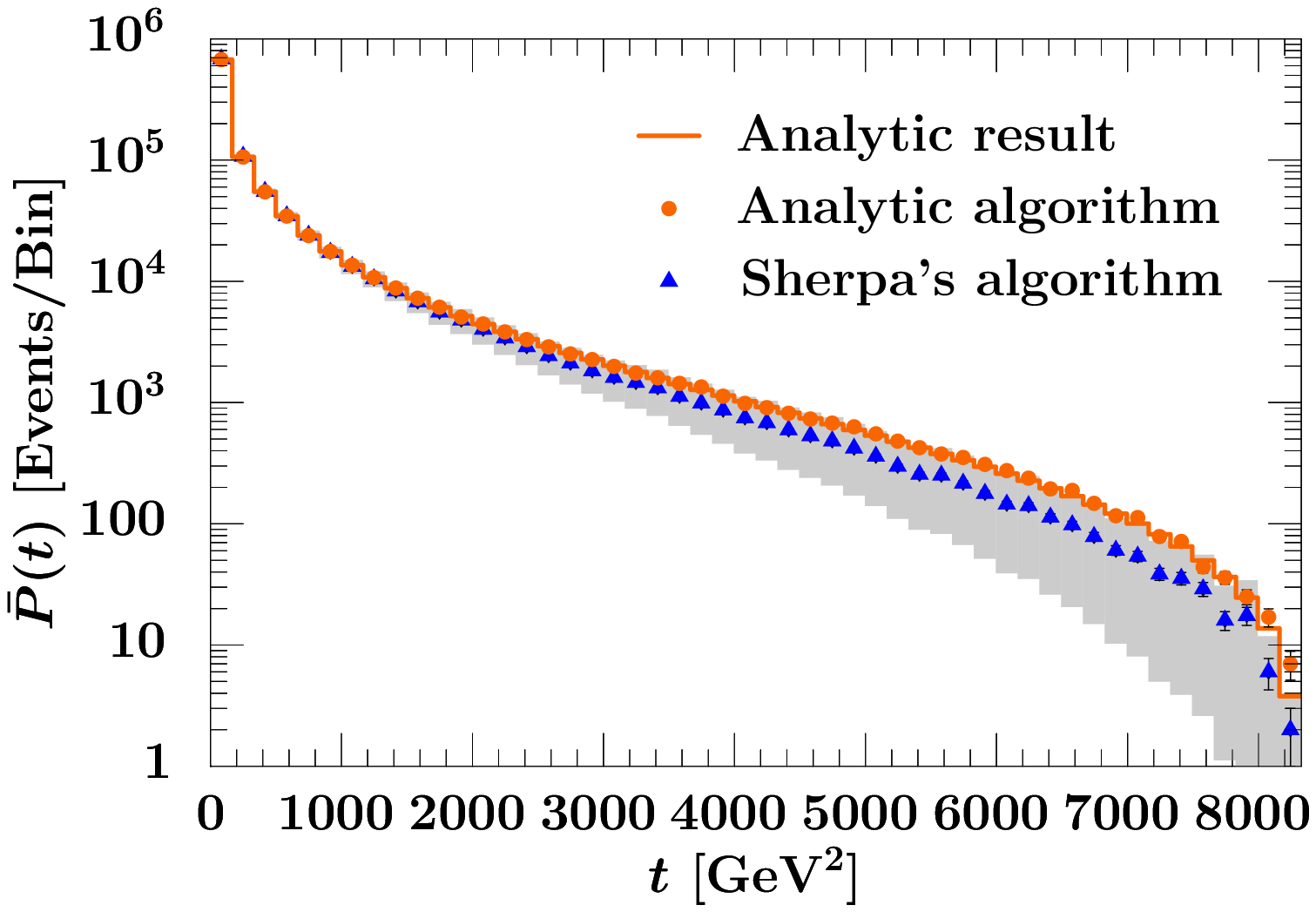}
\hspace{\columnsep}
\includegraphics[width=.45\textwidth]{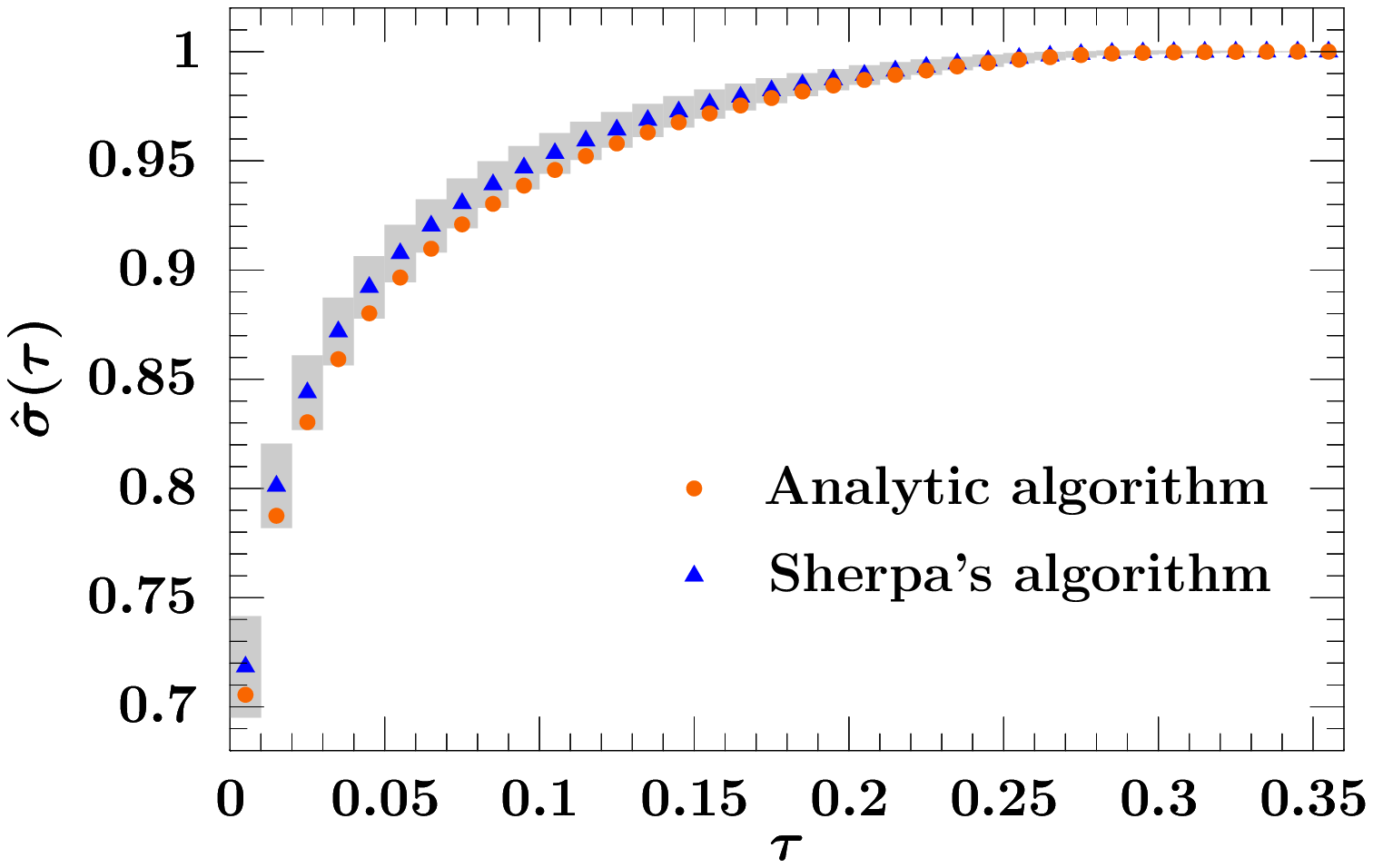}}
\caption{Comparison of the analytic parton shower [medium (orange) dots] and \sherpa's parton shower [dark (blue) triangles]. Left: The function $\bP(t)$ as defined in Eq.~\eqref{Pbar_def}. The solid (orange) line shows the result obtained from Eq.~\eqref{Pdouble}, and the gray shading gives an estimate of the expected size of power corrections. Right: The integrated thrust distribution, where the gray shading shows the expected size of finite perturbative corrections. The error bars show the statistical uncertainties.}
\label{fig:Pbar_thrust}
\end{figure*}

To illustrate the difference between the algorithms at the level of a single branching, we generated one million double branches starting from $t_\mo = (91.2\,\GeV)^2$ in the mother's rest frame, and look at the average of the double branch probability
%%%
\begin{align}
\label{Pbar_def}
\bP(t) &= \int \! \df t_L\, \df z_L\, \df t_R\, \df z_R \, \frac{\delta(t-t_L)+\delta(t-t_L)}{2}
\nn\\ & \qquad
 \times P(t_L,z_L;t_R,z_R)
\,.\end{align}
%%%
The results are shown on the left of Fig.~\ref{fig:Pbar_thrust} for the original \sherpa\ algorithm [dark (blue) triangles] and the analytic algorithm [medium (orange) dots]. The solid (orange) line shows the analytic result for $\bP(t)$ obtained from Eq.~\eqref{Pdouble}. As expected, it matches the distribution generated with the analytic algorithm. One can also see the suppression of the original algorithm at large $t$. The difference between the two algorithms for large $t$ is well within the expected size of power corrections Eq.~\eqref{DeltaP}, indicated by the gray shading.

To show the effect on fully showered events, we generated one million events with \sherpa\ using the original and the analytic algorithm. A typical physical observable is the thrust distribution, $\df\sigma/\df T$, which measures the jettiness of a given event, with $T \to 1$ for two narrow jets and $T \to 1/2$ for spherical events. On the right of Fig.~\ref{fig:Pbar_thrust} we show the integrated thrust distribution
%%%
\begin{equation}
\hat\sigma(\tau) \equiv \frac{1}{\sigma} \int_{1-\tau}^1\!\df T\, \frac{\df\sigma}{\df T}
\end{equation}
%%%
obtained with the original algorithm [dark (blue) triangles] and the analytic algorithm [medium (orange) dots]. Here, the gray shading gives an estimate of the size of perturbative corrections. As expected, compared to the original algorithm, the analytic algorithm suppresses small values of $\tau$, corresponding to branchings at small $t$, but within the estimated size of perturbative corrections.

\begin{figure}[t]
\centerline{\includegraphics[width=0.95\columnwidth]{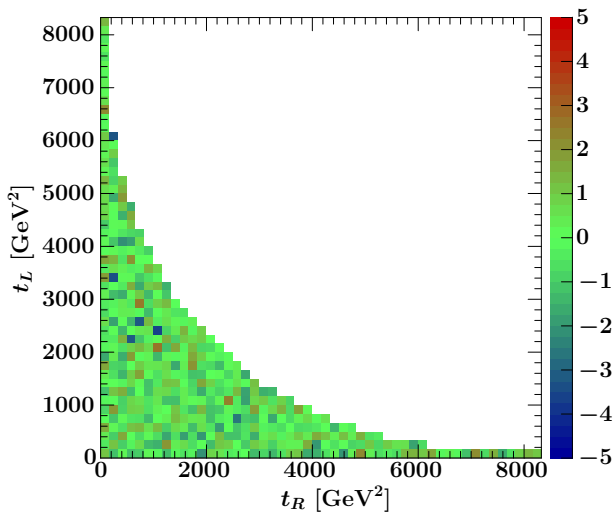}}
\caption{Pull distribution for the double branch probability $P(t_L,t_R)$ defined in Eq.~\eqref{Pdouble_t}. See text for further explanation.}
\label{fig:PtLRpull}
\end{figure}

As a further check that the analytic algorithm distributes according to the known probability Eq.~\eqref{Pdouble}, we keep the left and right branches separate and consider the double branch probability integrated over $z_{L,R}$
%%%
\begin{equation}
\label{Pdouble_t}
P(t_L,t_R) = \int \!dz_L\, dz_R\, P(t_L,z_L;t_R,z_R)
\,.\end{equation}
%%%
The result of the analytic algorithm agrees well within the statistical uncertainties with the analytic expectation from Eq.~\eqref{Pdouble}. This is illustrated in Fig.~\ref{fig:PtLRpull}, where we show the pull distribution for $P(t_L, t_R)$, i.e. the difference between the generated and analytic distributions divided by the statistical uncertainty of the generated distribution.

%%%%%%%%%%%%%%%%%%%%%%%%%%%%%%%%%%%%%%%%%%%%%%%%%%%%%%%%%%%%%%%%%%%%%%%%%%%%%%%%
\section{Reweighting events}
\label{sec:reweighting}
%%%%%%%%%%%%%%%%%%%%%%%%%%%%%%%%%%%%%%%%%%%%%%%%%%%%%%%%%%%%%%%%%%%%%%%%%%%%%%%%

We have shown how to construct an analytic parton shower algorithm which distributes events according to a known probability distribution $P_\PS(\{t_i,z_i\})$. This result opens up the possibility to obtain events that are distributed according to some other distribution $P_\new(\{t_i,z_i\})$ by proceeding in three steps:
\begin{enumerate}
\item Generate events using the analytic algorithm described in this work.
\item Assign the weight $P_\new(\{t_i,z_i\})/P_\PS(\{t_i,z_i\})$ to each event.
\item If desired, unweight the event sample by vetoing events according to their relative weights.
\end{enumerate}
The third step is optional and only needed if one desires final events with unit weight. Similarly, the allowable size of the weights will depend on the specific application. The power of this approach is twofold. First, it can provide a very efficient way to distribute according to some distribution $P_\new$, which may not be possible otherwise. Second, since the reweighting does not change the event kinematics, it can be applied at any later stage in the event generation, in particular, after the detector simulation. We like to stress again that, for this reweighting approach to be possible, it is essential to know the exact form of $P_\PS$, i.e., it would be insufficient to only know the leading terms in the power expansion of $P_\PS$. We now discuss two immediate applications.

%===============================================================================
\subsection{Distributing and Matching Matrix Elements}
%===============================================================================

We first consider the case where $P_\new = P_\ME$ is given by the differential distribution obtained from full matrix element calculations. Even though it is possible to obtain the squared amplitude, and thus $P_\ME$, for a moderate number of partons in the final state, it is still quite difficult to distribute events according to $P_\ME$. The reason is that $P_\ME$ has large peaks arising from poles in the amplitudes due to on-shell intermediate particles. This makes it impossible to employ a simple hit-or-miss algorithm with random numbers drawn from a flat distribution. Instead one has to rely on numerically inverting the integral of $P_\ME$ over phase space. Since the dimensionality of phase space increases with each additional parton, the required numerical integrations quickly become very time-consuming.

The parton shower describes the same IR physics as the full QCD amplitude. It thus contains the same poles from on-shell intermediate particles, and the resulting weights $P_\ME/P_\PS$ in step 2 are expected to be moderate. Generating events with a parton shower is several orders of magnitude faster than the numerical phase-space integrations required otherwise. Hence, it should be possible to accommodate relative weights even of $\ord(100)$ to $\ord(1000)$ and still achieve a reasonably high efficiency. As the events will later be run through a detector simulation, step 3 has to be included such that the final events have unit weight. One way to think of this is that the parton shower acts as a phase-space generator that automatically contains the correct pole structure of the matrix elements.

From this viewpoint, one could consider a much simpler version of the Markov Chain algorithm that only attempts to capture the underlying pole structure, without trying to address other important effects (e.g. getting the correct soft limit via the angular ordering) that are already included in the matrix elements. This would provide an alternative and efficient algorithm to distribute events according to known matrix element expressions.

Nevertheless, to obtain realistic, fully exclusive events one still has to attach a parton shower to the matrix element calculations, which is usually nontrivial due to double counting issues. There are a few dedicated algorithms~\cite{MLM,CKKW,Lonnblad:2001iq} with several implementations~\cite{Gleisberg:2003xi,Mrenna:2003if,Lavesson:2005xu} available to consistently match tree-level matrix elements for many partons with parton showers. In addition, several approaches are pursued by now to incorporate matrix elements for the first hard emission at next-to-leading order~\cite{Frixione:2002ik,Kramer:2003jk,Nason:2004rx,Latunde-Dada:2006gx}. To combine these two separate classes of matrix element corrections, some work has been carried out in Ref.~\cite{Nagy:2005aa}. In this respect, the SCET-based approach of Ref.~\cite{Bauer:2006qp} seems quite promising.

Using the analytic parton shower, one can generate fully showered events with many partons in the final state and then reweight the $n$ hardest emissions to the matrix element result for $n$ final-state partons. This idea is similar to the older merging method used by \pythia\ to correct the first shower emission~\cite{Bengtsson:1986hr,Miu:1998ut,Norrbin:2000uu,Sjostrand:2000wi}. In our case, one assigns to each event the weight $P_\ME(\{t_k,z_k\})/P_\PS(\{t_k,z_k\})$ in step 2, where $k=1,\ldots,n$ numbers the $n$ partons with the largest $t$. It follows that any observable sensitive to the distribution in the $n$ hardest partons and inclusive in all other partons will be determined by the full matrix element result, while any further emission is determined by the parton shower. In this way the analytic parton shower not only allows one to efficiently distribute matrix elements, but in addition provides a simple and powerful tool to match matrix elements and parton showers. An implementation of this result will be given elsewhere.
Note that the matching is completely determined at the analytic level by the form of $P_\ME$. In principle, it could be carried out for any $n$ and at any order in perturbation theory.

%===============================================================================
\subsection{Parton Shower Tuning and Uncertainties}
%===============================================================================

The reweighting can also be performed after the events have been run through a detector simulation, which is the most time-consuming part of the event generation. Hence, one can obtain sets of events with different underlying distributions, performing only a single run of the detector simulation. Of course, to make maximal use of the simulated events, one would now like to have $\ord(1)$ weights and also skip step 3.
Therefore, to generate events one would still use the best available matrix element calculations matched with the analytic parton shower, as described above.

One advantage of using the analytic parton shower is that one can update already simulated events at any later time to the newest theory or parameters. Furthermore, having analytic control over all parameters in the parton shower allows one to estimate uncertainties arising from input parameters like $\alpha_s(m_Z)$ or quark masses, and from higher-order power and perturbative corrections. In all cases $P_\new$ is simply given by $P_\PS$ computed with a different set of parameters. Moreover, one could study different scheme choices in the parton shower, provided of course one has analytic control over $P_\PS$ corresponding to the new scheme. For example, one could study the scale at which $\alpha_s$ is evaluated or even the choice of the evolution variable.

As a specific application, we look at power corrections and the tuning of the parton shower. Tuning a parton shower is crucial to get a good description of the experimental data, and one is, in effect, adjusting unknown power corrections to fit the data. The analytic control over the parton shower gives access to a simple and systematic way to tune it to data. We can introduce power corrections by adding nonsingular terms to the splitting function $f(t,z)$ entering Eq.~\eqref{Psingle_precise}. For example, for $f_{q\to qg}(t, z)$ as in Eq.~\eqref{fqqg},
%%%
\begin{equation}
f_\new(t,z) = \frac{\alpha_s C_F}{2\pi} \biggl[\frac{1}{t}\frac{1 + z^2}{1 - z} + g(t) \biggr]
\,,\end{equation}
%%%
where $g(t)$ is a nonsingular function of $t$ (in general, $g$ could also depend on $z$). This change will affect the branching probability for large $t$, but will leave branches with small $t$ nearly unaffected. Hence, adjusting $g(t)$ changes the power corrections included in the parton shower, which can be used to tune the parton shower.

\begin{figure}[t]
\centerline{\includegraphics[width=0.95\columnwidth]{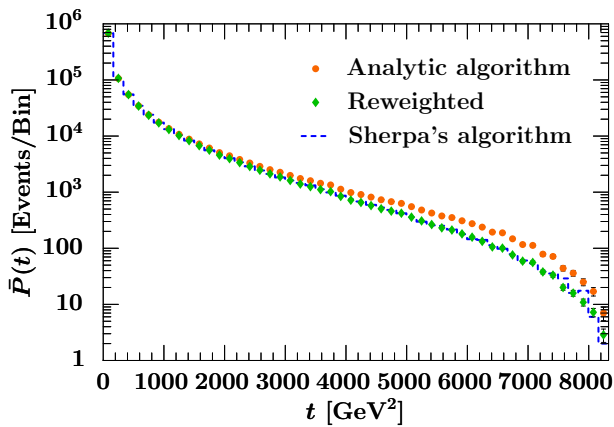}}
\caption{The function $\bP(t)$, as defined in Eq.~\eqref{Pbar_def}, for the analytic parton shower [medium (orange) dots], after reweighting [light (green) diamonds], and for \sherpa's parton shower [dashed (blue) line]. Error bars indicate statistical uncertainties. See text for further explanation.}
\label{fig:Pbar_reweighted}
\end{figure}

To illustrate this, we would like to adjust the analytic algorithm such that the average of its double branch probability, $\bP(t)$ [see Eq.~\eqref{Pbar_def}], roughly agrees with the original algorithm. We use the simplest possible power correction, $g(t) = a/t_\mx$. As discussed in Sec.~\ref{subsec:comparison}, the analytic algorithm enhances large $t$ compared to the original algorithm, so we need $a < 0$. In Fig.~\ref{fig:Pbar_reweighted} we show the result for $\bar P(t)$ from running the analytic algorithm [medium (orange) dots] and then reweighting each event to a new splitting function with $a = -1.5$ [light (green) diamonds], together with the result from running the original \sherpa\ algorithm [dashed (blue) line]. The reweighted result agrees remarkably well with the original algorithm, given the simple form of the power correction.

While this example only serves as an illustration, the analytic control over the tuning parameters is extremely useful. After the detector simulation, tuning the parton shower to data amounts to parametrizing the power corrections $g(t)$ in terms of a few parameters and fitting them to data. The uncertainties from power corrections can then be estimated by varying the tuning parameters in some range.

%%%%%%%%%%%%%%%%%%%%%%%%%%%%%%%%%%%%%%%%%%%%%%%%%%%%%%%%%%%%%%%%%%%%%%%%%%%%%%%%
\section{Conclusions}
\label{sec:conclusions}
%%%%%%%%%%%%%%%%%%%%%%%%%%%%%%%%%%%%%%%%%%%%%%%%%%%%%%%%%%%%%%%%%%%%%%%%%%%%%%%%

Event generators are indispensable tools to compare theory and experiment at collider experiments. While distributions with relatively few final-state partons can be computed directly from the matrix elements of the underlying theory, parton showers have to be employed to generate final states with a large number of partons. They rely on splitting functions, which are derived in the collinear or soft limit of the underlying theory. The conservation of four-momentum in each step of the algorithm, although technically a subleading effect, is an important ingredient to obtain realistic predictions from parton showers to compare to data.

In standard parton shower algorithms, the implementation of four-momentum conservation introduces a cross correlation between different branches, such that the final probability to produce a given event is not equal to the product of individual branching probabilities. In this work we have shown that a few simple modifications of an existing algorithm yield an analytic parton shower algorithm that conserves four-momentum at each vertex, but distributes according to a known analytic distribution. This makes it possible to generate events with exact knowledge of the probability with which each event was generated, and to reweight the events to a different distribution at any later stage in the event generation.

We have studied the specific case of a virtuality-ordered final-state parton shower. Considering the type of modifications, we expect the extension to a corresponding initial-state parton shower to be straightforward. It should also be possible to apply the same ideas to parton showers using different ordering variables.

The analytic parton shower proposed here in conjunction with the reweighting approach provides a powerful tool for experiment and theory, and in the last section we have given two examples of this. First, it facilitates the distribution of events according to full matrix elements, which is otherwise hindered by the need for time-consuming phase-space integrations, and at the same time provides a very generic way to match matrix elements with the parton shower. Second, generated events can be reweighted even after they have been run through a detector simulation. This allows one to update simulated events at any later time. It also greatly simplifies the study of higher-order corrections and parameter dependences in the parton shower as well as the estimation of uncertainties arising from these sources. For example, it provides a convenient way to tune the parton shower to data by simply fitting parameters characterizing power corrections to the data.

%%%%%%%%%%%%%%%%%%%%%%%%%%%%%%%%%%%%%%%%%%%%%%%%%%%%%%%%%%%%%%%%%%%%%%%%%%%%%%%%
\acknowledgments
%%%%%%%%%%%%%%%%%%%%%%%%%%%%%%%%%%%%%%%%%%%%%%%%%%%%%%%%%%%%%%%%%%%%%%%%%%%%%%%%

We are grateful to Johan Alwall, Peter Skands, and Kerstin Tackmann for helpful conversations.
This work was supported in part by the Director, Office of Science, Office
of High Energy Physics of the U.S.\ Department of Energy under the
Contract DE-AC02-05CH11231. CWB would also like to acknowledge support from the
DOE OJI program, and an LDRD from LBNL.

\end{document}